# Simulation of highly idealized, atomic scale MQCA logic circuits

Dmitri E. Nikonov, George I. Bourianoff, Paolo A. Gargini

Technology Strategy, Technology and Manufacturing Group, Intel Corp., SC1-05, 2200 Mission College Blvd., Santa Clara, California 95052, USA

*Abstract* – Spintronics logic devices based on majority gates formed by atomic-level arrangements of spins in the crystal lattice is considered. The dynamics of switching is modeled by time-dependent solution of the density-matrix equation with relaxation. The devices are shown to satisfy requirements for logic. Switching speed and dissipated energy are calculated and compared with electronic transistors. The simulations show that for the highly idealized case assumed here, it is possible to trade off size for speed and achieve lower power operation than ultimately scaled CMOS devices.

## I. INTRODUCTION

The International Technology Roadmap for Silicon (ITRS) [1] and many other publications have articulated the need for alternative logic devices to continue Moore's law [2] scaling beyond that of conventional CMOS devices and ferromagnetic, spin based devices [3] are frequently mentioned as one of those possible alternative technologies. One of the frequently cited motivations for considering spin based logic technologies is the possibility of achieving low power operation relative to ultimately scaled CMOS. However, to the best of the authors' knowledge, there are no theoretical comparisons of the properties of ultimately scaled MQCA logic circuits relative to ultimately scaled CMOS circuits. Here we look at ultimate limit of magnetic QCA [4], where the cells are formed by single spins of electrons localized in the crystal lattice. We describe a possible realization in perovskite structure of manganites, in Section II. Since we deal with adjacent crystal lattice cells, the majority gates look like crossed wires, and in this sense they resemble another type of spintronic devices, based on motion of domain walls [5].

The suitable formalism to simulate the dynamics of single spins are the quantum microscopic equations (density matrix equations) [6], rather than the phenomenological Landau-Lifshits-Gilbert equations [7]. The interactions between spins can be modeled by the Heisenberg Hamiltonian [8], see Section III for more detail. In the majority of works, the steady-state solution is investigated for these systems [9]. Typically Monte Carlo simulation methods are used for this purpose [10]. Only a few works are devoted to time-dependent processes, e.g. [11] and even those only treat simple geometries. Our goal here is to apply this formalism to spintronic devices, which necessitates the study of time-dependent switching in more complex layouts of circuits.

In Section IV, we show that spintronic devices fulfill the requirements for logic and thus can actually form viable logic circuits. One of the paramount requirements is power gain for the signal, which has previously been modeled for magnetic QCA [12]. Theoretical schemes promise lower dissipation in spintronics [13], mainly due to operation out of thermal equilibrium. In this paper we do not consider this type of operation, at the end of each switching event the devices reaches thermal equilibrium. In Section V, we calculate physical parameters of spintronic device switching and compare them with conventional electronic circuits implemented with end of the roadmap CMOS devices. It is shown that subject to the many assumptions made in the analysis, spintronic devices are slower but more power efficient. We argue that such spintronic devices and circuits are therefore worthy of further research in an effort to understand the ultimate limits imposed by implementation in practical circuits. If the practical engineering and fabrication issues could be successfully resolved, such spin based logic circuits could compliment CMOS in certain applications but would not constitute a general purpose CMOS replacement.

## II. MANGANITES AND MAJORITY GATE LOGIC

Majority gate logic (or QCA) [14] is made up of wires containing cells, each holding a certain state of the computational variable. This variable can be spin polarization, magnetization, etc. [4]. Logical '1' corresponds to orientation along a certain preferred axis, and logical '0' to the orientation opposite to this axis. An inverter performs the logical NOT function – its output is opposite to the input. A majority gate has three inputs and one output (Figure 1), which is determined by the majority vote of the computational variables of the inputs. The function of the majority gate contains both logical AND and OR function of any two of its outputs, depending on whether the third output is set to '0' or to '1'. Thus the majority gate enables reconfigurable logic.

In order to allow the up-stream cells influence the downstream cells, all cells except the input ones need to be transferred to the intermediate state, e.g. direction of spin perpendicular to the preferred axis. The difference should be that the intermediate state is meta-stable and the logical '0' and '1' should be stable states corresponding to global minima of energy in the absence of the external field. This can be achieved for example by choosing the logical states along the easy axis of the ferromagnetic dot. In other words, anisotropy between the hard and easy axes is crucial for operation of spin QCA. The process of bringing all cells into the intermediate state before each computation is called "clocking" [15]. It can be implemented for example by applying a global magnetic field, which would align all the spins along a hard axis. At the end of the computation, the spins in each cell are is "tipped" into one of the two stable directions ( the logical '0' or '1' by the perturbing influence of the adjacent cells.

We would like to push this concept to the extreme, in which the QCA cell correspond to a unit cell in a crystal lattice and spins of individual electrons store the computational state, Figure 1. This arrangement is possible in the perovskite crystal structure of manganites, e.g. LSMO, $La_{1-x}Sr_xMnO_3$. For the review of manganite physics see [16,17]. The manganese ion is positioned in the center of a cube of the lattice; oxygen is at the faces of the cube and lanthanum and strontium ions are at the vertices of the cube. Manganese has electrons on its d-shell. The three electrons in the $t_{2g}$ subshell are localized and have the combined spin S=3/2. The electrons in the de-localized $e_g$ subshell interact with nearest neighbor manganese atoms via the process of double-exchange mediated by surrounding oxygen ions. This results in ferromagnetic alignment of localized spins.

A simple scheme of the spintronic majority gate device [18] is shown in Figure 2. We envision that a few, and ultimately a single monolayer of a complex metal oxide (e.g. LSMO) can be grown on a substrate (which can be e.g. STO). This film is patterned into wires of the majority gate, each only a few crystal lattice cells wide. We appreciate the tremendous challenges of patterning such thin wires, but leave them outside the scope of this paper. The upside of using the crystal lattice cells is that they are virtually identical. This solves the problem of non-uniformity of cells which plagued the performance of macroscopic QCAs.

The interface with electronic circuits is implemented as ferromagnetic contacts over the wires separated by a tunneling barrier from them. The input state of magnetization in wires is set by spin torque [19] originating from a current across the tunneling barrier. The output direction of magnetization is measured by the effect of tunneling magnetoresistance [20]. We also envision the option of ferroelectric electrodes (e.g. out of PZT) which can modulate or switch the ferromagnetic state in LSMO wires. All these interfaces (as well as the computation in QCA) do not require application of magnetic fields (unlike magnetic RAM, for example). The only instance of application of magnetic field is clocking, to align magnetic moments between the computing cycles. This field needs to exceed the coercivity field of the ferromagnet (LSMO) and is easily achievable. Even then it is a global magnetic field applied to the whole chip, rather than local magnetic fields (as in MRAM). Therefore the problems with scaling the size of circuit elements caused by magnetic field cross talk and the problem of high power dissipation from currents generating local magnetic fields do not arise in spintronic logic considered here.

## III. QUANTUM DYNAMICS OF MAJORITY GATES

We employ a simple model of interacting spins on a lattice. We are using the model to explain the operation of spintronic QCA. Though it is only a crude approximation to the very rich physics of manganties [17], it captures the features we care about – ferromagnetic alignment of spins and anisotropy. Further in this paper, for the sake of simplicity we perform the calculations with spin S=½ rather than spin S=3/2. The results are qualitatively similar. We use the well-studied Hamiltonian of the non-isotropic Heisenberg model [8]

$$H = -\sum_i G\mathbf{S}_i \cdot \mathbf{B} - \sum_{ij}\left(J_x S_i^x S_j^x + J_y S_i^y S_j^y + J_z S_i^z S_j^z\right)$$

(1)

where $S_{x,y,z}$ are the operators of projections of the spin vector, $J_{x,y,z}$ are the coupling constants, $G = gq/(2m_e)$ is the gyromagnetic constant, $\mathbf{B}$ is the effective magnetic field. In simulations considered below, the magnetic field will correspond to the clocking global magnetic field. Also setting of the input magnetizations in the wires by spin torque will be modeled as the strong bias magnetic field which keeps the direction of input magnetic moments essentially fixed. The indices i and j correspond to lattice sites and the second sum in (1) includes only pairs of nearest neighbors. The value of the effective Heisenberg coupling is chosen to give the correct Curie temperature $T_C \approx 350K$ for La$_{0.7}$Sr$_{0.3}$MnO$_3$, and according to [21]

$$J = \frac{3k_B T_C}{zS(S+1)} \approx 30 meV$$

(2)

An extra factor of 2 in our case is caused by the difference of definition of the constant. Here $z = 4$ is the number of nearest neighbors. Note that its value would be different if we modeled a more realistic case with S=3/2. It is in order-of-magnitude agreement with the estimate [22] from the hopping integral $t \approx 0.4 eV$ and the on-site charging energy $U \approx 6 eV$

$$J = \frac{2t^2}{U} \approx 53 meV .$$

(3)

With this Hamiltonian, we can solve the equation for the density matrix $\rho$ [6] for the spins at each site.

$$\frac{d\rho}{dt} = \frac{-i}{\hbar}[H, \rho] - \gamma\left(\rho - f(\langle H \rangle)\right),$$

(4)

where $f$ is the equilibrium Fermi distribution with the ambient temperature $T$, and we assume a very fast spin relaxation rate $\gamma \approx 10^{14} s^{-1}$ due to strong coupling of spin and orbital state characteristic for complex metal oxides. The expectation values of physical quantities are calculated as quantum averages of corresponding operators $O$:

$$\langle O \rangle = Tr[O\rho]$$

(5)

These equations contain terms corresponding to the relaxation arising from contact with the external thermal reservoir such as phonons, lattice defects etc. Thus they account for dissipative processes, but are not capturing thermal fluctuations explicitly. In their treatment of spin interaction this approach is equivalent to mean-field theory, but in its dynamics, rather than in its static form.

As we noticed in Section II, the correct operation of QCA is ensured by the anisotropy. It may arise from the inherent anisotropy of the material or can be induced by stress from deposition on a substrate with a different lattice constant. The most significant effect in wires is usually the demagnetization – change of the effective magnetic field **B** compared to the external magnetic field $\mathbf{B}_{ext}$ due to the fringing fields of the ferromagnet.

$$\mathbf{B} = \mathbf{B}_{ext} - \mu_0 \overline{\mathbf{N}} \cdot \mathbf{M}$$

(6)

It is described by the demagnetization tensor $\overline{\mathbf{N}}$ and favors magnetization along the wire. In this paper we consider the case of an easy x-axis. It is not an equivalent of demagnetization, but allows us to simplify the geometry, treatment, and interpretation. We assume $J_y = 30 meV$, $J_x = 1.1 J_y$, and $J_z = 1.1 J_y$.

The general features of the dynamics of spin systems described by this model can be foreseen based on the nature of the terms in Eqs. (1) and (4). The magnetic field will tend to align magnetic moments along its direction. The Heisenberg coupling term favors ferromagnetic (parallel) alignment of neighboring spins. Therefore the magnetization next to fixed input magnetic moments will evolve to be parallel to them. Under the influence of a few neighboring spins, a spin will tend to choose the direction determined by their net sum. Anisotropy will favor alignment along the easy axis (x-axis on this case). Domains of spins aligned close to parallel will be separated by domain walls. The change of spin direction will propagate as a spin wave from the fixed spins to other parts of the circuit. The time that is required to reach the steady state is determined by the slowest among the inverse relaxation rate and the time for the spin wave to traverse the circuit. The plots in this paper show the initial and the final state for switching. The animations of these simulations can be downloaded from [23]. They confirm the expected features of the spin system's evolution.

With our model, we demonstrate the functioning of the elements of the majority gate logic. It should be noted, that realistic majority gate circuits may be more complicated in order to ensure synchronous operation and to exclude spontaneous switching. We do not discuss these aspects here and rather focus on a simple model which is adequate for estimates of the device scaling. In subsequent plots, to designate the direction of magnetization in the plane of the chip, we use a color scheme conventional for micromagnetics: red – right, blue – up, green – left, yellow – down. In the initial state all magnetic moments are aligned up by application of the external magnetic field. The exception is the edge magnetic moments, which are used as input to the system and set by external circuits using spin torque transfer. To initiate the computation, the magnetic field is removed, and spin magnetic moments align to minimize their interaction energy.

Propagation of magnetization in the wire is presented in Figure 3. The leftmost magnetic moments are set to point to the right (logical '1'). All the rest of the wire is aligned up by the clocking field. Spins next to the input quickly evolve to align to be parallel due to spin relaxation. The spin at the front of the wave of turning spin is pointing at an angle to the wire and causes spins to the spins ahead to precess to point down. As the spin wave propagates along the wire it settles into a pattern of a standing spin wave – regions of magnetization along or opposite to the wire separated by domain walls. Its period changes depending on the strength of coupling and geometry of the wire, but remains fixed along the wire. If the wire length is close to odd number of half periods, its rightmost spin ends up opposite to the input one. Therefore the wire of a carefully chosen length can serve as an inverter.

The operation of the majority gate is shown in Figure 4. The inputs are specified by edge magnetic moments of the top, left and bottom wires. In the case pictured they are '0', '1', and '1', or left, right and right, up and up, respectively. The spin waves travel from the inputs to the center of the cross, causing the alignment in the direction of input spins. The three waves collide at the center and compete for to determine alignment of spins there. Once the steady state is reached, a spin wave continues along the output wire. This wave carries the direction of magnetization reached in the center of the majority gate. For our set of parameters, the state of the output is '1' (magnetization to the right), as required by majority voting. The top wire forms a domain pointing left and remains separated from the rest of the gate by a domain wall. However, if the relaxation rate is significantly slower, there are cases when magnetization in the center overshoots in its oscillations and ends up pointing left (logical '0'). This regime would break down the majority voting and should be avoided.

Thereby our simple quantum dynamics simulation confirms the designed operation of the majority gate devices as needed to perform expected logical functions.

## IV. VIABILITY OF SPINTRONIC LOGIC

As mentioned in Sections I, CMOS transistors have been tremendously successful as elements of logic circuits. A set of unique properties of field effect transistors enabled this success. Nowadays, CMOS devices are proposed and their merits are analyzed without regard of whether they can perform logic operation. In fact a device needs to satisfy multiple requirements to serve as a logic element. The first set of them are the indispensable requirements ("tenets") [24] without which a device would be useless for logic:

1. Non-linear characteristics (related to the signal-to-noise ratio)
2. Power amplification (gain>1)
3. Concatenability (output of one device can drive another)
4. Feedback prevention (output does not affect input)
5. Complete set of Boolean operators (NOT, AND, OR, or equivalent)

Secondly, the device needs to be characterized in terms of its physical parameters. Ones specifically called out ITRS [1] are.

1. Size (connected to scalability)
2. Switching speed
3. Switching energy (connected to power dissipation)

Finally there are practical requirements promoting technological success of the new computing technology, which according to ITRS [1] are

1. Room temperature operation
2. Low sensitivity to parameters (e.g. fabrication variations)
3. Operational reliability

4. CMOS architectural compatibility (interface, connection scheme)
5. CMOS process compatibility (fabricated on the same wafer)

Though they are very important, their determination requires extensive research and development. Spintronics is in an earlier stage, and therefore these requirements will not be addressed in this paper. Our simulations justify that the considered device paradigm satisfies the tenets for logic.

Ferromagnetic state of the material ensures that the absolute value of magnetization is approximately constant outside the domain walls. The logic state in our devices is thus encoded by the angle of magnetization. In fact it is this angle (relative to the fixed magnetization of ferromagnetic electrodes) is what is measured in the tunneling magnetoresistance detection. The nonlinearity relative to the angle ensures that angle variations in one stage will not accumulate with variations in the second stage, but will instead be suppressed. We have performed a set of simulations of the switching dynamics of a majority gate. In these the two inputs were held constant at '0' and '1'. The angle of the remaining input was varied from 0 to $\pi$, which corresponds to change from '1' to '0'. The angle of the magnetization in the output wire is plotted in Figure 5. We see that it is strongly nonlinear and reminiscent of the CMOS inverter characteristic. By the same token, the slope in the middle point is related to gain. It is ~16 in the case shown.

Power gain is observed as the flux of energy from a spin to its neighbor increases as the domain wall propagates along the wire [15].

Concatenability of majority gate devices is obvious from the simulations. Maximum magnetization is uniform over the circuit and is determined by the properties of the material. The angle of magnetization varies from 0 to $\pi$ for either input of output.

In order to demonstrate isolation, we performed another set of simulations. After the steady state has been reach in previous simulations (as in Figure 4), we record this state and remove fixed inputs. Instead the angle of the rightmost magnetic moments is set, e.g. using spin torque. The variation of the average magnetization in the left wire is plotted in Figure 6. One can see that the change in that angle is small, $\sim 10^{-3}$. The reason for this small change is the anisotropy, which prevents the spins from switching if their direction is away from the metastable state. The practical significance of it is related to the fact, that even if the spin state in the next stage changes, it does not propagate back to the previous stage of the circuit. Therefore clocked majority gates maintain good isolation of input from output.

Presence of the complete set of the logical operations has been discussed above, in Section II.

Therefore the presented majority gate devices satisfy all tenets of logic.

## V. PHYSICAL PARAMETERS: COMPARISON OF SPINTRONICS AND ELECTRONICS

Finally we characterize the spintronic devices in terms of physical parameters – length, speed, energy. Another set of simulation is performed as described in Section III, now with various widths of the wires. Gate length is equal to the length of the cross in the majority gate, 5 times the width of the wire. For each case we plot density of projection of spin on each axis and the electron density (which remains constant) in the output piece of the majority gate, as in Figure 7. We see that the spin projections evolve in an oscillatory manner and then settle to their steady state value due to relaxation. The switching time is approximately taken as the time of evolution from 10% to 90% of the range between the initial and the final state among all spin projections. We also calculated the dissipated power by determining the quantum average of the Hamiltonian multiplied by the second (dissipative) term in the right-hand side of Eq.(4) and summed it over all spins in the majority gate. Switching energy is obtained as the integral over time of the dissipated power, as in Figure 8. From that curve we see that majority of energy is dissipated in the short interval after the clocking field is removed. It is the energy associated with a large angle difference between the input spins and their neighbor spins.

We compare the obtained values with the device parameters for CMOS transistors: the channel length of the transistor $L$, the intrinsic switching time $\tau = CV/I_{on}$ and the switching energy $E = CV^2$, where $C$ is the capacitance of the gate, $V$ is the voltage swing, and $I_{on}$ is the maximum current when the transistor is on. The comparison is done in accordance with the benchmarking methodology of Ref. [25]. The CMOS dataset contains historic data up to year 2003 [26] and projections for the future up to year 2016 according to ITRS [1]. The years when corresponding generations of CMOS transistors went into production are labeled on the plots.

We observe that the structure of atomic-level majority gates permits scaling to 2nm, beyond the sizes envisioned by the semiconductor industry roadmap. One of the advantages of spintronics at such small sizes is a fixed number of electrons in a device, while in a CMOS transistor gate it is not fixed. Therefore spintronic circuits will not experience the shot noise which would plague electronic circuits.

The switching delays of spintronic devices (Figure 9) are found to be longer for the same length than in electronic transistors. This is explained by the fact that the speed of domain walls is about an order of magnitude smaller than typical injection velocities in semiconductors.

Switching energies in spintronic devices (Figure 10) are found to be smaller than in electronic transistors. We see it as the main advantage of spintronics: it allows the trade-off of speed for energy. From this we derive the energy-delay product (Figure 11). Surprisingly one can see that spintronic devices fall on essentially the same scaling line as electronic devices. Spintronic devices scale further to approach the quantum mechanical limit $E\tau \geq \pi\hbar$, according to the Margolus-Levitin theorem [27].

In order to compare the computational efficiency of various technologies, it is necessary to compare the power dissipation and computational throughput on a *per chip basis*. To do that, we need to relate the power dissipation per chip to the power dissipation per device and similarly adjust for the number of device switching times per clock cycle. Based on data in the ITRS [1] we estimate that only 0.1% of the chip area will be made up of transistors, the individual transistor will switch in 0.5% of the clock cycle, only 3% of the transistors will switch at any one clock cycle and only1.6% of the total energy dissipation of the chip comes from dynamic switching energy of the transistors. These observations for CMOS may be converted to an alternative set of phenomenological factors which will be defined in a similar manner for other technologies

- average chip area per device in units of gate length squared, $C_L = 1000$,
- ratio of the clock period to the intrinsic device switching time, $C_t = 200$,
- activity factor, the fraction of transistors switched in any clock cycle, $C_{ac} = 0.03$,
- ratio of total energy dissipated per switching a transistor (with the account of interconnects and clocking) to the intrinsic switching energy, $C_E = 60$.

Therefore the density of logic elements is

$$D_L = \frac{1}{C_L L^2},$$

(7)

computational throughput per unit area is

$$N_c = \frac{C_{ac} D_L}{C_t \tau},$$

(8)

and power dissipation per unit area is

$$P_d = C_E E N_c.$$

(9)

Since we assumed these factors to be the same for both electronic and spintronics devices (which is uncertain at the moment), they preserve the relative values, but give realistic numbers for power dissipation and computational throughput encountered in logical circuits. Note that in the present simulations, even the smallest gate contains spins of tens of electrons, unlike the model of [28] which deals with a single electron. The following advantage of spintronic circuits exists despite this many-particle nature.

We see in Figure 12, that spintronic devices provide a few times smaller power dissipation for the same throughput at the end of their scaling than the electronic devices projected by ITRS. Though the simulated spintronics devices are somewhat slower, they have higher densities, and therefore comparable values of computational throughput. Since switching energy in spintronic devices is smaller, power dissipation in spintronic circuits is smaller too for the same value of computational throughput.

## VI. Conclusion

In summary, we are describing a paradigm of spintronic devices based on majority gates formed by spins of electrons localized in crystal lattice cells. These devices embody logic devices scaled to their ultimate size. We have applied the formalism of quantum dynamics to simulate the switching of the logic elements – the inverter and the majority gate. We calculate the switching speed and energy dissipated in the process. We demonstrate that these devices satisfy necessary tenets to serve as a part of a functional logic circuit. Considered spintronic circuits require application of a global magnetic field only for initialization before the computing cycle ("clocking"). The device physical parameters – length, speed, and dissipated energy – are compared with those of realistic CMOS transistors. It is found that, though at a slower switching speed, spintronic devices dissipate smaller energy per switching. This fact gives them advantage in further increase of density of logic devices on chip, according to the Moore's law. Spintronic devices exhibit several times lower dissipated power per unit area of the chip for the same computational throughput.

# References

[1] Semiconductor Industry Association, "International Roadmap for Semiconductors", 2007. Available: http://public.itrs.net/.

[2] G. E. Moore, "Cramming more components onto integrated circuits," *Electronics*, April 19, pp. 114-117, 1965.

[3] I. Zutic, J. Fabian, S. Das Sarma, "Spintronics: Fundamentals and Applications", Rev. Mod. Phys., v. 76, 323-410, 2004.

[4] A. Imre, G. Csaba, L. Ji, A. Orlov, G. H. Bernstein, and W. Porod, "Majority Logic Gate for Magnetic Quantum-Dot Cellular Automata", Science, v. 311, pp. 205-208, 2006.

[5] D. A. Allwood, G. Xiong, C. C. Faulkner, D. Atkinson, D. Petit, and R. P. Cowburn, "Magnetic Domain-Wall Logic", Science, v. 309, pp. 1688-92, 2005.

[6] M. O. Scully and M. S. Zubairy, "Quantum Optics", Cambridge University Press, Cambrdige, United Kingdom, 1997.

[7] T. L. Gilbert, "A Lagrangian Formulation of the Gyromagnetic Equation of the Magnetization Field," Phys. Rev. 100, 1243 (1955).

[8] G. D. Mahan, "Many-Particle Physics", 3$^{rd}$ ed., Kluwer, New York, 2000.

[9] D. P. Landau and M. Krech, "Spin dynamics simulations of classical ferro- and antiferromagnetic model systems: comparison with theory and experiment", J. Phys.: Condens. Matter, v. 11, pp. R179–R213 (1999).

[10] H de Raedt and A. Lagendijk , "Monte Carlo Simulation of Quantum Statistical lattice Models", Phys. Rep., v. 127, pp. 233-307 (1985).

[11] S. W. Lovesey and E. Balcari, "A theory of the time-dependent properties of Heisenberg spin chains at infinite temperature", J. Phys.: Condens. Matter, v. 6, pp. 1253-1260 (1994).

[12] J. Timler and C. S. Lent, "Power gain and dissipation in quantum-dot cellular automata", J. Appl. Phys., v. 91, pp. 823-831, 2002.

[13] D. E. Nikonov, G. I. Bourianoff, P. A. Gargini, "Power dissipation in spintronic devices away from thermodynamic equilibrium", J. Superconductivity and Novel Magnetism, v. 19, # 6, p. 497-513, 2006.

[14] C. S. Lent and P. D. Tougaw, "A Device Architecture for Computing with Quantum Dots", Proc. IEEE, v. 85, pp. 541-557, 1997.

[15] G. Csaba, A. Imre, G. H. Bernstein, W. Porod, and V. Metlushko "Nanocomputing by Field-Coupled Nanomagnets", IEEE Trans. Nanotechnology, v. 4, pp. 209-213, 2002.

[16] Y. Tokura, "Correlated-Electron Physics in Transition-Metal Oxides", Physics Today, July 2003, pp. 50-55.

[17] E. Dagotto, "Nanoscale Phase Separation and Colossal Magnetoresistance", Springer, Berlin, 2003.

[18] G. I. Bourianoff, D. E. Nikonov, and J. F. Zheng, US Patent No. 7,212,026, 2007.

[19] Katine, J. A., F. J. Albert, R. A. Buhrman, E. B. Myers, and D.Ralph, 2000, ''Current-driven magnetization reversal and spin-wave excitations in Co/Cu/Co pillars,'' Phys. Rev. Lett. **84**, 3149–3152.

[20] Julliere, M., 1975, ''Tunneling between ferromagnetic films,'' Phys. Lett. **54A**, 225–226.

[21] Blundell S, "Magnetism in Condensed Matter", Oxford Press, Oxford, 2001.

[22] D. Khomskii, "Electronic Structure, Exchange and Magnetism in Oxides", pp. 89-116, in "Spin Electronics", Eds. M.J. Thornton, M. Ziese, Lecture Notes in Physics, vol. 569, Springer, Berlin, 2001.

[23] Animations of simulations. Available www.nanohub.org.

[24] "Nanoelectronics and Information Technology", ed. R. Waser, Wiley-VCH, Weinheim, Germany, 2003.

[25] R. Chau, S. Datta, M. Doczy, B. Doyle, B. Jin, J. Kavalieros, A. Majumdar, M. Metz, and M. Radosavljevic, "Benchmarking Nanotechnology for High-Performance and Low-Power Logic Transistor Applications", IEEE Trans. Nanotech., v. 4, pp. 153-158, 2005.

[26] R. Chau, private communication, 2005.

[27] N. Margolus and L. B Levitin, "The maximum speed of dynamical evolution", Physica D, vol. 120, pp. 188-195, 1998.

[28] Zhirnov VV, Cavin RK, Hutchby JA, Bourianoff GI," Limits to binary logic switch scaling - a gedanken model", Proceedings of the IEEE, 91, 11, 1934–9, 2003.

**Figures**

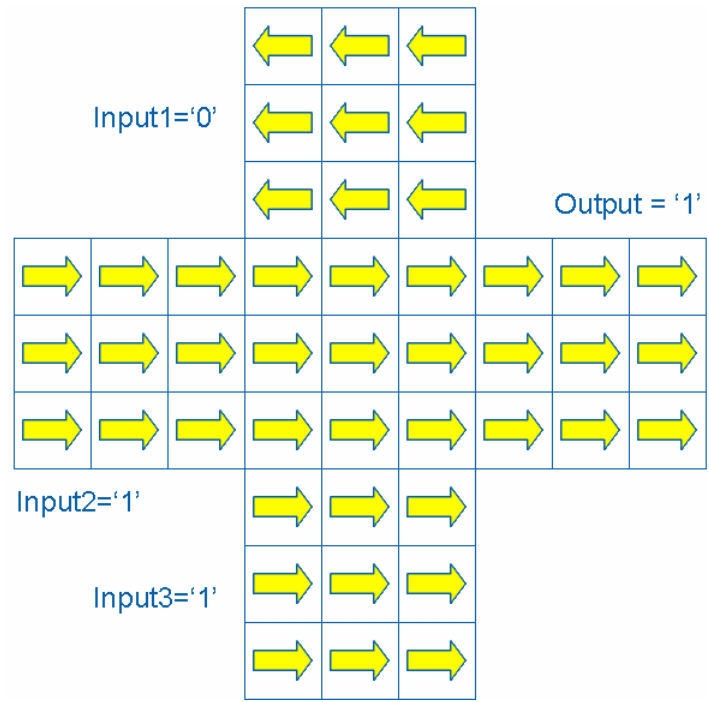

Fig. 1. (color online) A top-view scheme of a majority gate with wires 3 lattice cell wide. The easy axis is assumed in left-right direction. Directions of spins in the cells are approximate.

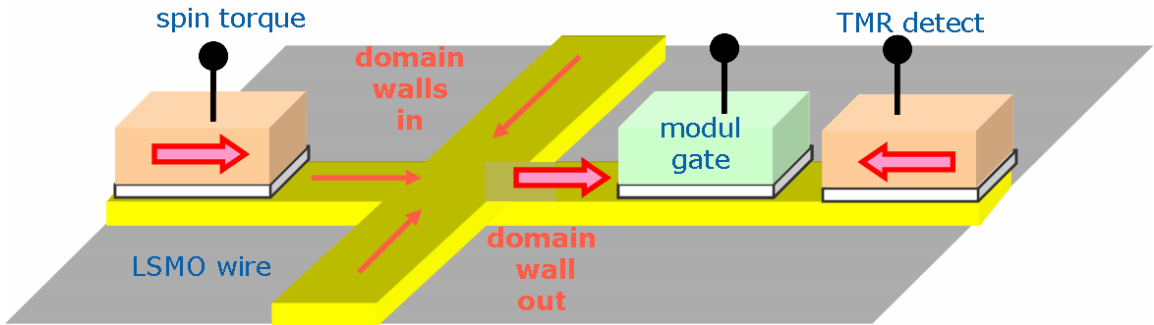

Fig. 2. (color online) Envisioned device structure for spintronic majority gates. Monolayer LSMO wires are formed on a substrate (e.g. STO). Input directions of magnetizations are set by ferromagnetic electrodes, e.g. via spin torque switching. Option to switch or modulate magnetization with a gate (e.g. made of PZT). The direction of the output magnetization is detected via tunneling magnetoresistance effect.

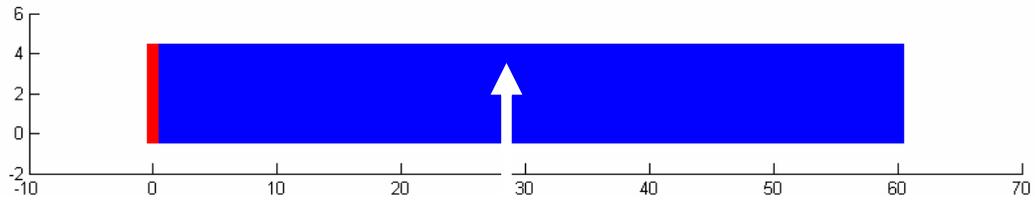

a)

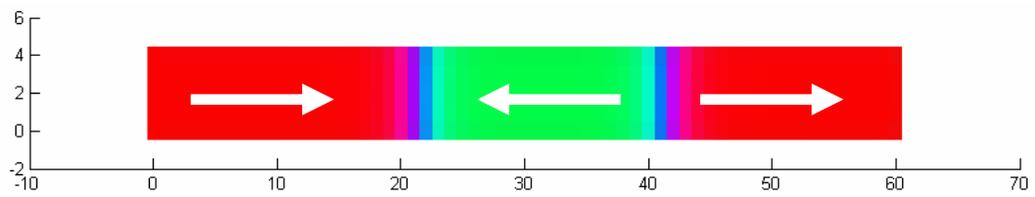

b)

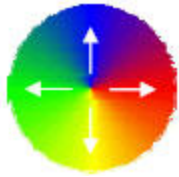

c)

Fig. 3. (color online). Magnetization state in a straight wire (top view), in the beginning of switching (a) and the end of switching (b). Color scheme for direction of magnetization (c).

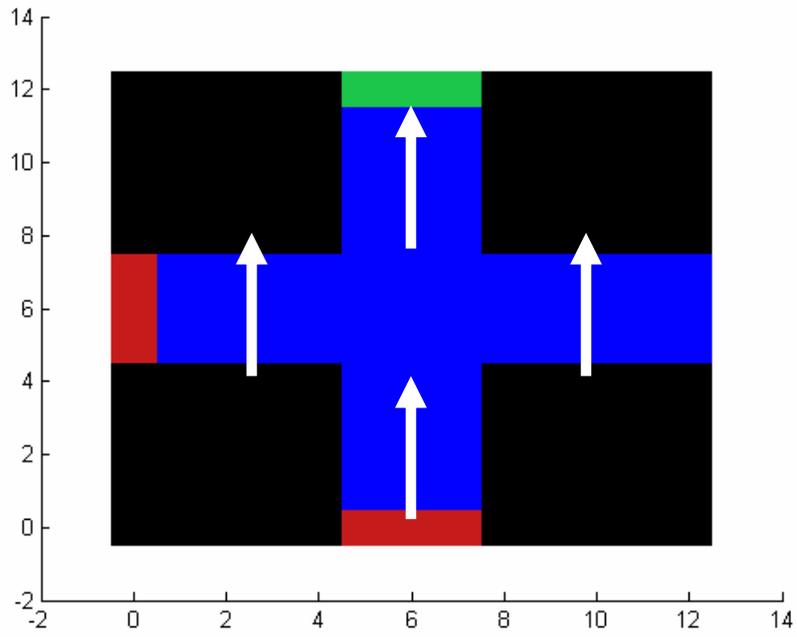

a)

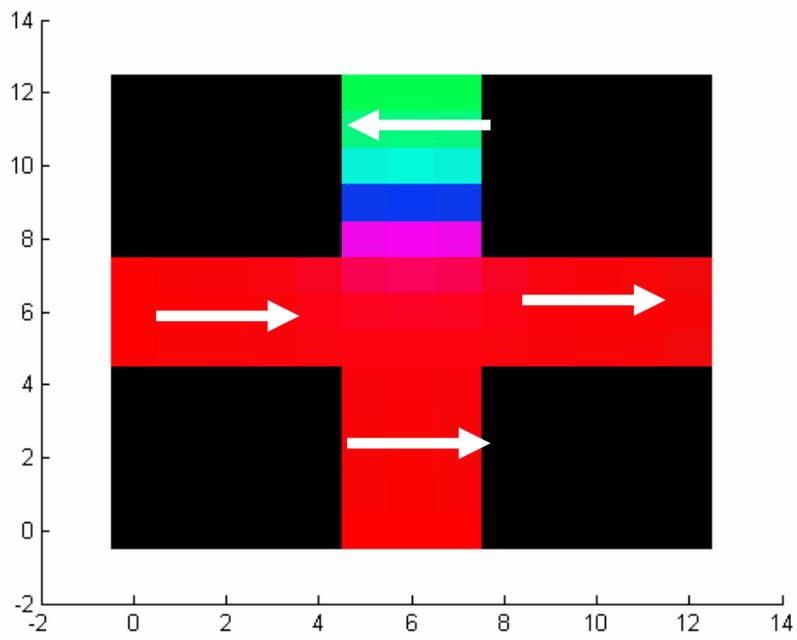

(b)

Fig. 4. (color online) Magnetization state in a majority gate (top view), in the beginning of switching (a) and the end of switching (b). Case of input '0', '1', '1' shown.

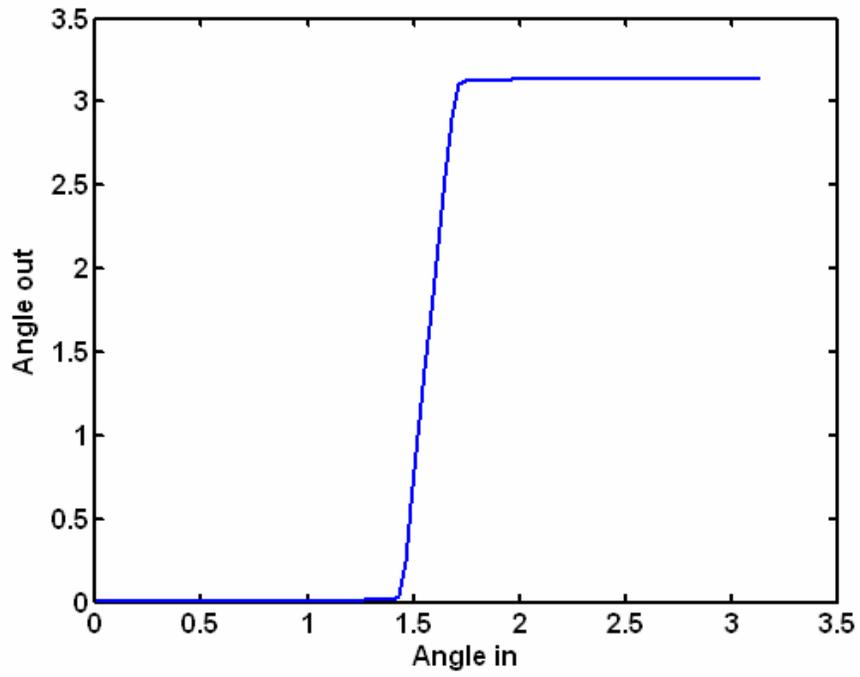

Fig. 5. Nonlinear characteristic of a magnetic gate: angle of the output magnetization vs. angle of the magnetization in input2, while input1='0', and input3='1'.

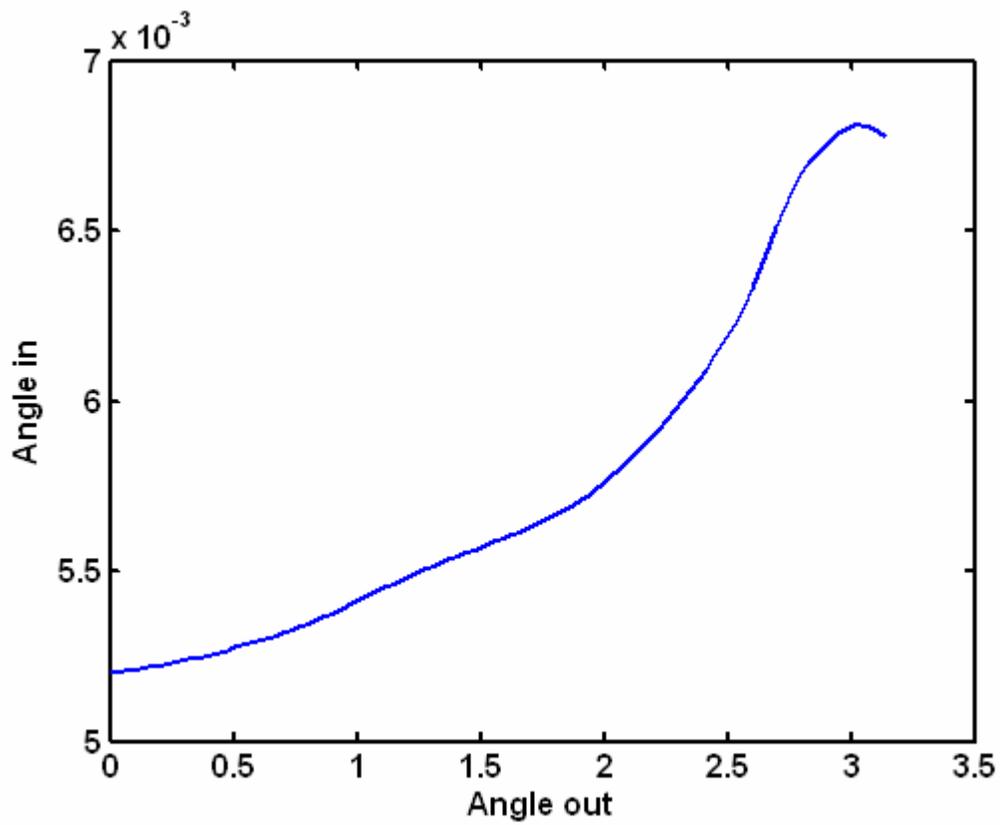

Fig. 6. Input-output isolation characteristic: change of the angle in input2 when the magnetization angle in the output is varied after switching in the case of Fig 4.

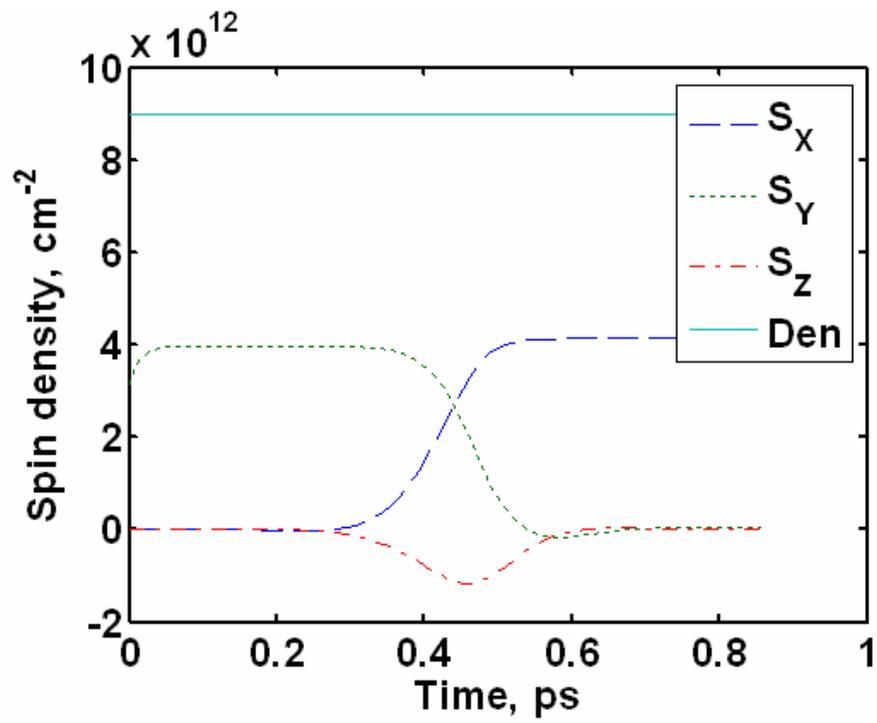

Fig. 7. (color online) Time dependence of the spin state in the output wire in the case of Fig. 5. Sx, Sy, Sz are densities of the spin projection on axes (with -1 factor corresponding to electrons), Den is the electron density (constant).

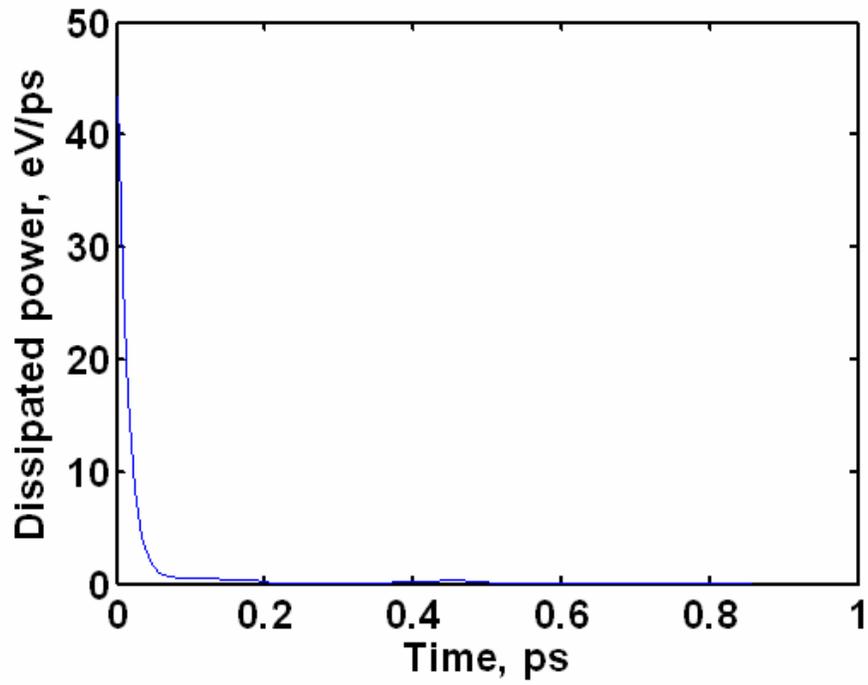

Fig. 8. Time dependence of the dissipated power in the whole majority gate for the case of Fig. 5. Total dissipated energy 0.80eV.

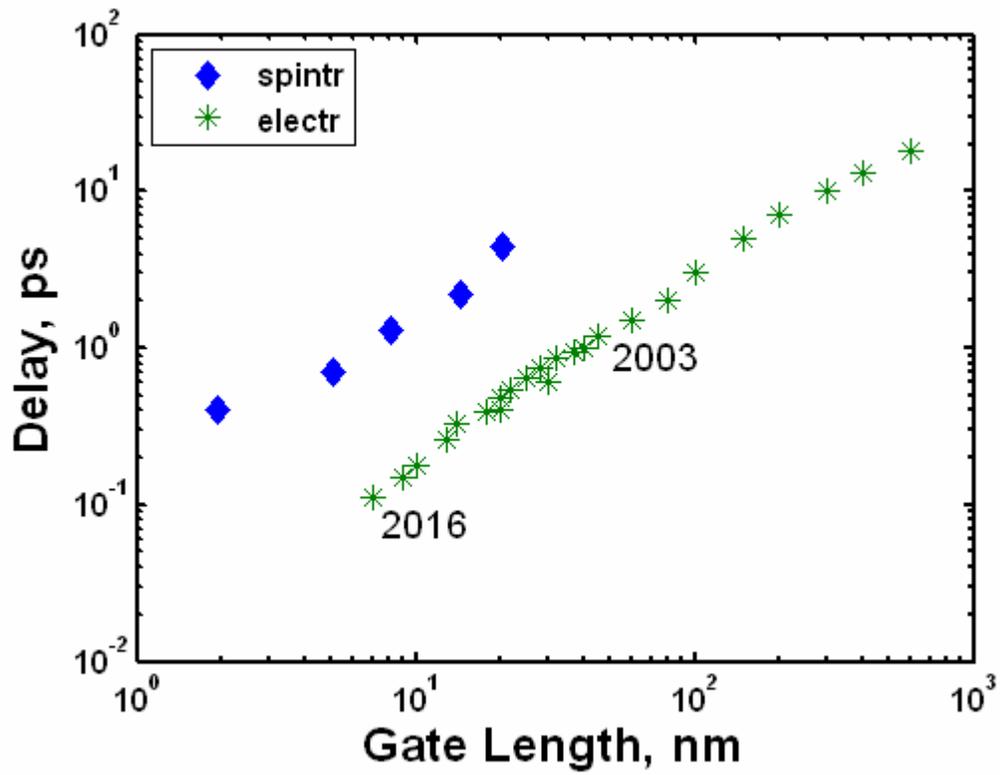

Fig. 9. Switching delay vs. gate length. Stars – CMOS transistors; diamonds – spintronic majority gates. Year of production labeled for CMOS.

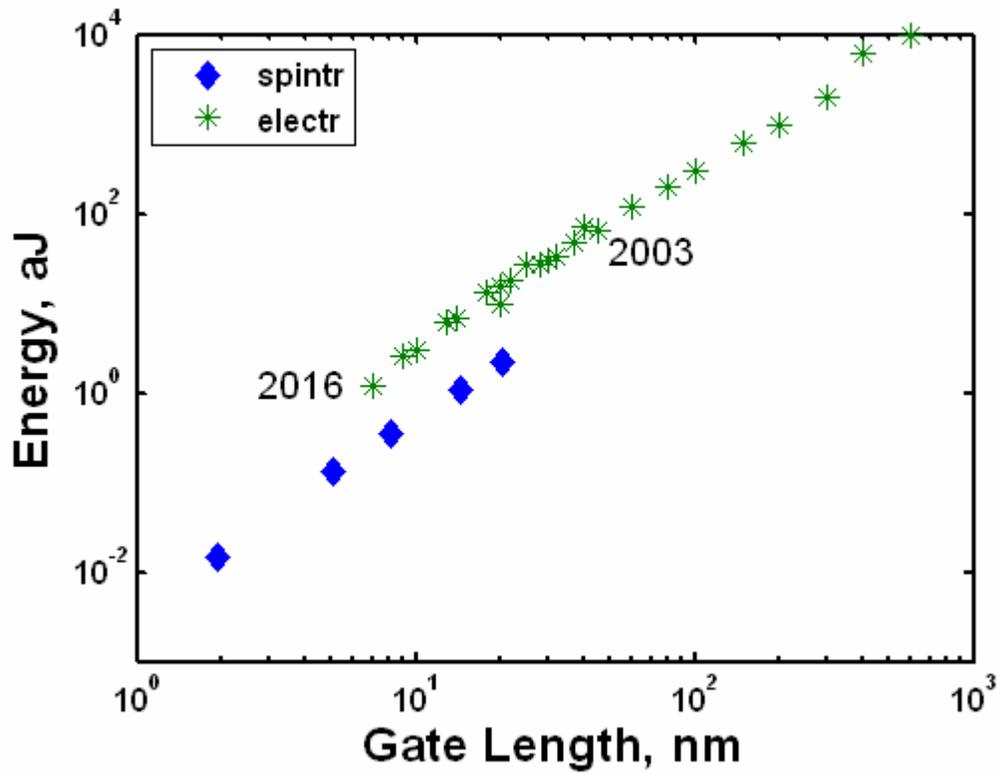

Fig. 10. Switching energy vs. gate length; similar to Fig. 9.

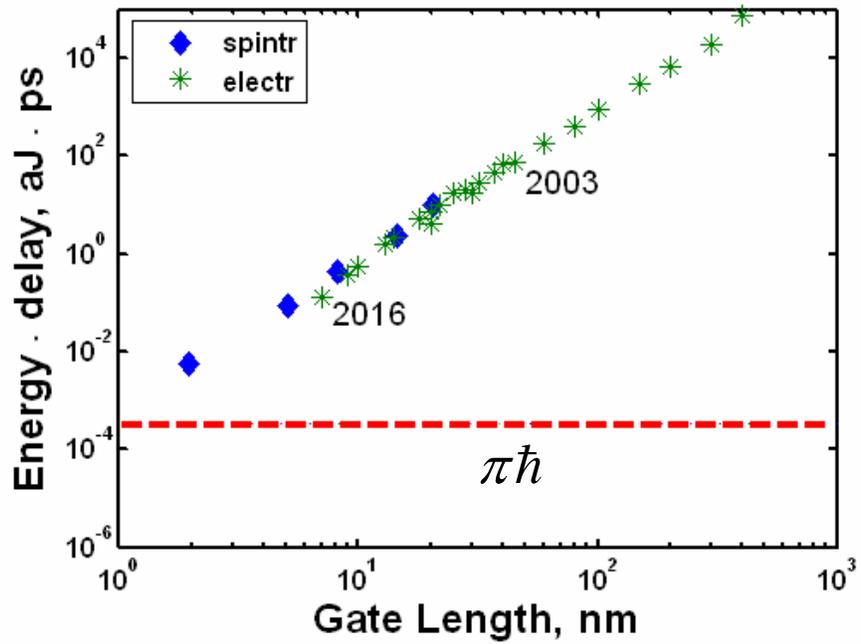

Fig. 11. Energy-delay product; similar to Fig. 9. Dashed line marks the quantum limit.

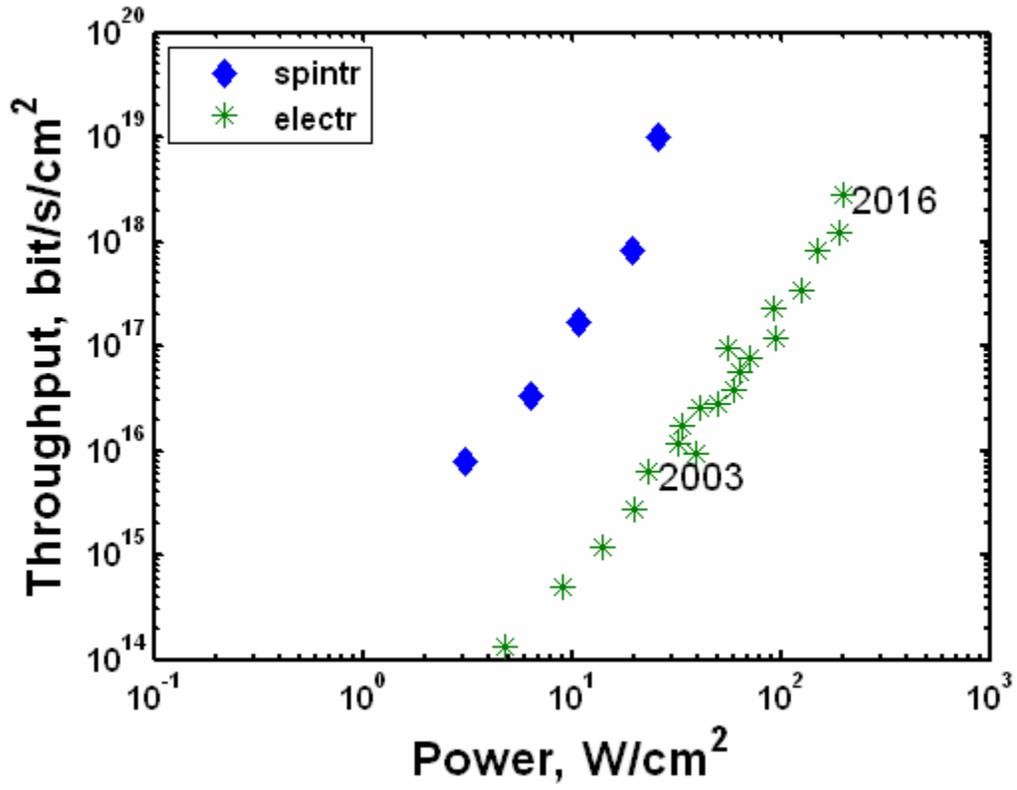

Fig. 12. Computational throughput vs. dissipated power, both per unit area; similar to Fig. 9.